\documentclass[prl,twocolumn,showpacs,amsmath,amssymb]{revtex4}
\usepackage{graphicx}
\usepackage{dcolumn}
\usepackage{bm}
\begin{document}
\title{Floating phase  in a  dissipative Josephson junction array}
\author{Sumanta Tewari and John Toner}
\affiliation{Department of Physics and Institute of Theoretical
Science, University of Oregon, Eugene, OR 97403}
\author{Sudip Chakravarty}
\affiliation{Department of Physics and Astronomy, University of
California Los Angeles, Los Angeles, CA 90095-1547}
  \date{\today}
\begin{abstract}
We consider dissipative quantum phase transitions in Josephson
junction arrays and show that the disordered phase in this
extended system can be viewed as an unusual floating phase in
which the states of local $(0+1)$-dimensional elements (single
Josephson junctions) can slide past each other despite arbitrary
range spatial couplings among them. The
unusual character of the metal-superconductor quantum critical
point can be tested by measurements of the current voltage
characteristic. This  may be the simplest and most natural example
of a floating phase.
\end{abstract}
\pacs{} \maketitle

It has recently been recognized \cite{Toner1} that
phases and phase transitions in a given dimension can be
embedded in a higher dimensional manifold. One realization of this is in
a  stack of two-dimensional (2D) layers of coupled $XY$ spins with
interlayer couplings along  the layer-normals \cite{Toner1}. For a
suitably chosen (and fairly complicated)
set of competing interlayer couplings involving higher order gradients,
it was  shown in Ref. \cite{Toner1} that the 3D-system can exhibit
2D-behavior: power-law ordered phases in each layer below a
characteristic temperature, $T_{KT}$ , and a high-temperature
disordered phase with exponentially decaying correlations. The transition
separating the two phases remains 2D Kosterlitz-Thouless type. Below
$T_{KT}$, the 2D power-law phases in each layer `slide' past one another
(a ``floating'' phase) \cite{Toner1}. Strikingly, such a phase is
impossible for more conventional Josephson-type couplings between
the layers. Another interesting case is the recently discussed
quantum theory of the smectic metal state in stripe phases
\cite{Emery}, which also involves gradient couplings.
Unfortunately, there are no definitive experimental realizations
of a floating phase, although there is some evidence for it in
lipid-DNA complexes \cite{DNA} in which lipid bilayers are able to
slide over each other without cost in energy \cite{Safinya}.

In this Letter we show that such a floating phase naturally occurs in
the ground state of a system of coupled resistively shunted
Josephson junctions \cite{Goldman}.
The floating phase in this
case is a state where the ground states of $(0+1)$-dimensional
elements (single Josephson junctions) can slide past each other
despite coupling between them.
Experimentally, this phase will behave like
a metal with finite resistance\cite{Chakravarty0,Fisherdiss}.
The striking point is that no additional, higher order
gradient couplings need be invoked,
in contrast to the stack of XY spins treated in Ref.~\cite{Toner1}.

The existence of such a floating phase in this system has been
conjectured before\cite{Chakravarty0,CIKZ,Fisher}.
The arguments presented there are, however, incomplete, since further
than nearest-neighbor interactions, which are always present in any
real system, are not considered in these treatments. This is particularly
worrisome since the treatment of XY stacks\cite{Toner1} found many
regions of the phase diagram in which such interactions destroyed the
floating phase.
In this paper, we show that no such problem arises here: at least for
sufficiently small further neighbor interactions, the floating phase is
robust.

 A concrete realization of longer ranged Josephson couplings
relevant to the present model is a system of superconducting grains embedded in a
metallic medium. The Josephson coupling $J_{ij}$ between the
grains at $r_{i}$ and  $r_{j}$ falls off very slowly at $T=0$,
specifically $J_{ij}\propto \frac{1}{|r_{i}-r_{j}|^{2}}$ in two
dimensions \cite{Larkin};  this is due to pair field-pair field
correlation function in a normal  metal.

\begin{figure}[htb]
\includegraphics[scale=0.4]{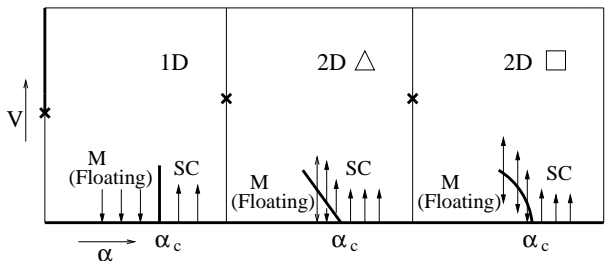}
\caption{Phase diagram and perturbative RG flows for 1D, 2D
triangular lattice (3D face centered cubic), and 2D square lattice (3D simple cubic). M and
SC indicate metal and superconductor, respectively. The thick
lines correspond to line of fixed points. Crosses indicate special
points on the $\alpha=0$ axis denoting ($D+1$)-dimensional $XY$
transition. } \label{fig:flows}
\end{figure}

The prior work also argued\cite{Chakravarty0,Fisher,CIKZ} that
there will be a phase transition as the shunt resistance is
varied, between the floating phase and a ($D+1$)-dimensional
superconducting phase, in which the phases of the junctions are
coherently coupled. In addition to confirming this, we show from a
renormalization group analysis that this transition, for suitable
shunt resistance and small nearest neighbor Josephson coupling
$V$, is governed by a line of quantum critical points with
continuously varying critical exponents, which in dimensions
higher than one is very rare indeed. As we discuss below, the
upper critical dimension in this problem is infinity. The phase
diagram is shown in   Fig.~\ref{fig:flows}. In this figure,
$\alpha\equiv R/R_Q$, where $R$ is the shunt resistance, and
$R_Q\equiv h/4e^2$ is the quantum of resistance.

A similar phase diagram has been  seen  in a recent set of experiments
\cite{Takahide}, but the existence of continuously varying critical exponents is not yet established.
We emphasize, however, that the floating phase
is a metal, not an insulator: the metallic shunts  short the
current when the phase difference between the junctions
become incoherent. The
metallic behavior was not established in this experiment, but
only the insulating behavior of the unshunted array. Clearly, further
experiments  at sufficiently low temperatures will be necessary in
establishing the true nature of the floating phase. We expect that the current-voltage characteristic in
the floating phase to be a non-universal power law controlled by
dissipation $\alpha$, similar to the results obtained previously
\cite{Kane}. In the superconducting state, the
temperature-dependent non-universal power law will be  similar to
the classical XY model due to vortex unbinding in the presence of
an imposed current \cite{Kadin}.

As mentioned earlier,  the phase transition is very strange,
exhibiting continuously varying, non-universal critical exponents
that depend on where the phase boundary is crossed. Additionally,
the universality class of the transition depends {\it not at all }
on the dimensionality of the system, but only on the local
topology of the lattice.  In particular, for a one-dimensional
chain, the phase boundary  in the  $V$-$\alpha$ plane is vertical
for small $V$ (that is, independent of  $V$), while for {\it any}
lattice of {\it any} dimensionality that has closed loops of 3
nearest neighbor bonds (e.g., a 2D triangular or a 3D FCC
lattice),   the critical Josphson coupling $V_c$   obeys, for
$\alpha$ near $\alpha_c$  :
\begin{equation}
V_c(\alpha) = A(\alpha_c - \alpha) , \label{vctriang}
  \end{equation}
where $A$ is a non-universal constant.
For {\it any} lattice of
{\it any} dimensionality in which the smallest  closed
loops \cite{nodub} have {\it more} than  3 nearest neighbor bonds (e.g., a
2D square or 3D simple cubic lattice), the
critical Josphson coupling
$V_c$   obeys, for
$\alpha$ near
$\alpha_c$  :
\begin{equation}
V_c(\alpha) =\sqrt{A'(\alpha_c - \alpha)}   , \label{vcsquare}
  \end{equation}
where $A'$ is another non-universal constant.
We will hereafter refer to these  distinct lattice types as
``triangular-type" and ``square-type" respectively.

One non-universal critical exponent that can easily be measured
experimentally is the thermal exponent $\nu$, which controls the scaling
of characteristic temperatures $T_{ch}$ with distance from the $T=0$
quantum phase transition. One such characteristic temperature
for $\alpha$'s and $V$'s that lie in the superconducting
phase at $T=0$ is superconducting to normal transition temperature $T_c$.
For  $\alpha$'s and $V$'s that lie in the floating
phase at $T=0$, a characteristic temperature is the temperature at which
the resistivity of the array as a whole has its minimum\cite{Fisherdiss}.

We
predict that
\begin{equation}
T_{ch} \propto(\delta V)^\nu , \label{vcsquare}
  \end{equation}
with $\delta V \equiv |V - V_c|$ . With $\alpha_*\ne\alpha_c$ the
value of $\alpha$ at which the transition occurs,
  the non-universal exponent $\nu(\alpha_*)$ is given by
\begin{equation}
\nu(\alpha_*) = \frac{1}{A''(\alpha_c - \alpha_*)} , \label{vcsquare}
  \end{equation}
with $A''=1(2)$ for ``triangular(square)-type"  lattices.

The array of the Josephson-junctions coupled to local Ohmic
dissipation is described by the following
imaginary-time ($\tau$) action:
\begin{eqnarray}
{\cal{S}}/\hbar&=&\int_{0}^{\beta}[\frac{1}{2E_{0}}\sum_i(\frac{\partial
\theta_i}{\partial
\tau})^{2} d\tau + V\sum_{\langle
i,j\rangle}[1-\cos\Delta\theta_{ij}(\tau)]]d\tau \nonumber\\
   &+& \frac{\alpha}{4 \pi}\sum_{\langle
i,j\rangle}\frac{1}{\beta}\sum_{n}|\omega_{n}||\Delta
  \tilde{\theta}_{ij}(\omega_{n})|^{2 }+ {\cal{S}}_J/\hbar \label{action1}
  \end{eqnarray}
Here the sum $\langle i,j \rangle$ is over nearest neighbor pairs
in a lattice of arbitrary dimensionality and type.
$\Delta \tilde{\theta}_{ij}(\omega_{n})$ is a Fourier component of
the phase difference between the two superconducting grains,
$\Delta\theta_{ij}(\tau)$, with $\omega_n=\pm{2\pi n/\beta}$:
$n$ is an integer, and $\beta$ is the inverse-temperature. The  charging energy
of a grain, $E_{0}$, is
given by $E_{0}\propto 1/C$, where $C$ is the capacitance of the grain.
The  term containing $\alpha$
arises from integrating out the degrees of
freedom of the harmonic-oscillator bath with a linear spectral-function
(Ohmic dissipation)
\cite{Caldeira}.
An Ohmic shunt, by definition, transfers charge
continuously, while in quasiparticle tunneling in a superconductor
the charge is transferred in discrete quantized amounts
\cite{Schmid2}.

The term, ${\cal{S}}_J$, describes inter-junction couplings
mediated by longer-range Josephson-interactions among the grains,
which has not been considered previously.
These couplings are described by the
general action,
  \begin{equation}
{\cal{S}}_J/\hbar=-\sum_{[s_i]}\sum_{i}\int_0^{\beta}d\tau
J([s_i])\cos[\sum_j s_j\theta ({\vec r}_i + {\vec \delta}_j,
\tau)] \label{action2}
  \end{equation}
Here, $\theta ({\vec r}_i, \tau) $'s are the phases of the order
parameters in the  superconducting grains, the ${\vec \delta}_i$'s
are arbitrary separation vectors on the lattice (to allow for
arbitrary ranged couplings) and $s_i$ is an integer-valued
function of the layer number $i$ satisfying $\sum_i s_i = 0$. This
last constraint follows from the absence of an external field,
which implies ``rotation invariance" under adding the same
constant to all of the $\theta$'s. Note that the special case $s_0
=+1 , {\vec \delta}_0 = {\vec 0}, s_1 =-1 , {\vec \delta}_1 =
{\vec a_{\gamma}}$, where ${\vec a_{\gamma}}$ is a nearest
neighbor vector, is just the nearest neighbor Josephson coupling
$V$ in Eq.~(\ref{action1}).

To establish the existence of a dissipation-tuned quantum phase
transition, we use  a renormalization group analysis  that is
perturbative in $V$ and $J$. We divide the field $\theta_i(\tau)$
into slow  and fast components $\theta_{is}(\tau)$ and
$\theta_{if}(\tau)$ and write the partition function $Z$ as
  \begin{equation}
Z=Z_{0} \int\prod_i{\cal D}\theta_{is}(\tau)\exp{\left[-{\cal
S}_{0}^s/\hbar + \ln\langle e^{-{\cal
S}^{\prime}/\hbar}\rangle_{0f}\right]}  \label{partition}
  \end{equation}
Here $Z_{0}$ is a normalization constant, ${\cal S}_{0}^{s}$ is
the slow-frequency component of the quadratic part  of the action
containing the first and the third terms in Eq.~(\ref{action1}),
${\cal S}^{\prime}$ contains the remaining two terms, and
$\langle\ldots\rangle_{0f}$ denotes the average with respect to
the fast components of the quadratic part. The average in
Eq.~(\ref{partition}) contains a product of two factors involving
$V$ and $J$.

To leading order in $V$ and $J$, the coupling constants do not
mix. After computing the averages, we rescale $\tau$,
$\tau'=\tau/b$, $b$ being the scale factor, to restore the
original frequency cut-off, and then redefine the coupling
constants to complete the renormalization.  The dissipation term
in Eq.~(\ref{action1}) is dimensionless, and so it is held fixed
by the RG. The first term then has $\tau$-dimension $-1$. Thus, it is
irrelevant in the RG sense, and will henceforth be dropped for the
remainder of this paper\cite{dangerous}.

Writing  $b=e^l$, where $l>0$ is infinitesimal, and $z_0 = z/2$,
where $z$ is the coordination number of the
lattice, we have to one loop order \cite{CIKZ}:
\begin{equation}
\frac{dV}{dl}=(1-\frac{1}{z_0\alpha})V \label{Vrg}
  \end{equation}
In the absence of the $J$'s, the $\alpha$-axis with $V=0$ is a
line of stable fixed points for $\alpha<1/z_0$ and a line of
unstable fixed points for $\alpha>1/z_0$. The physical
interpretation of the stable fixed points is as follows: for
$\alpha<1/z_0$, the barrier $V$ among the different potential
minima of the variable $\Delta\theta_{ij}$ is irrelevant, and the
system described by $\Delta\theta_{ij}$ exhibits local quantum
fluctuations.

This local criticality can be seen from the equal-time correlation
function of the phase differences:
  \begin{equation}
  \langle \exp[i q \Delta\theta_{ij}(\tau)]\rangle
\propto  1/(\beta \hbar \omega_c)^{\eta ( q )} , \label{corr0}
\end{equation}
with $\eta(q)\equiv^{\text min}_{n\in Z}\eta'(q-n)^2 $, $ \eta '
\equiv 1/( z_0 \alpha)$, $q$ a real number, and
$\hbar\omega_c=E_0$. Note that the simplest such correlation
function one could imagine, namely $q = 1$, does {\sl not} decay
algebraically because $\eta(1) = 0$.  This result for $\eta ( q )$ is
familiar for surface roughening ~\cite{rough, srough}.
A similar
calculation shows that the unequal  time correlation functions of
the bond variables are also critical:
\begin{equation}
  \langle
\exp[iq(\Delta\theta_{ij}(\tau)-\Delta\theta_{ij}(0))]\rangle
=\frac{1}{(\omega_c\tau )^{2 \eta (q )}}
  \label{corr1} .
\end{equation}
Although the correlation functions of the bond variables
algebraically decay with time in any dimension, the same is not
true for the correlation of the site variables. We have computed
the unequal time site correlation functions \cite{future} and have
found that they vanish as $\tau\to \infty$; the power law form is
obtained for only $D\ge 3$. The locality of this phase follows
from the result that \cite{future}, at zero
  temperature, spatial
correlation functions of the grains,
$<\exp[iq(\theta_i(\tau)-\theta_{i+r}(\tau))]>$, are identically
zero in any dimension; spatially separated grains are completely
uncorrelated. Each junction is a dynamical
  system on its own, and there are only short-ranged correlations
between spatially separated junctions.

For $\alpha>1/z_0$, on the other hand, the barrier $V$ grows, and
the individual quantum systems saturate in one of the potential
minima. $V$ is like a local field term for the `bond-spin'
variable which becomes relevant above $\alpha=1/z_0$, and orders
the individual bond-spins. The transition at $\alpha=1/z_0$ is purely
local.

What happens
when longer-ranged spatial couplings among them are introduced?
Calculations precisely analogous to that leading to
Eq.~(\ref{Vrg}) give, to first order in $J$,
\begin{equation}
\frac{dJ}{dl}=[1-\frac{\Gamma([s_i])}{\alpha}]J, \label{Jrg}
\end{equation}
  with
\begin{equation}
\Gamma([s_i])\equiv \sum_{i,j }  s_i s_j U( {\vec \delta}_i -
{\vec \delta}_j). \label{Gammadef}
\end{equation}
The ``potential'' $ U( {\vec r}) \equiv \frac{1}{N}\sum_{{\vec
q}}(e^{i{\vec q} \cdot {\vec r}} - 1)/f_{\vec q}$, with $f_{\vec
q} \equiv  \sum_{{\gamma}}(1 - e^{i{\vec q} \cdot {\vec a_\gamma}}
)$, where the sum on $\gamma$ is over all nearest neighbors. It is
straightforward to show that $U({\vec r})$
  is nothing but the ``lattice Coulomb potential''
  of a unit negative charge at the origin, with the zero of
  the potential set at ${\vec r} = {\vec 0}$. That is, $U({\vec r})$ satisfies
  the ``lattice Poisson equation":
  $ \sum_\gamma U({\vec r} - {\vec a}_\gamma) - z U ({\vec r}) = -
\delta_{{\vec r}, {\vec 0}}$.
  For a symmetrical (e.g., square, hexagonal, cubic) lattice, the
left-hand side is
  just the ``lattice Laplacian'', approaching $\nabla^2 U \times O(a^2)$ where
  ($a \equiv |{\vec a}_\gamma |$).

The quantity $\Gamma$ in (\ref{Jrg}) and (\ref{Gammadef}) is then
clearly just equal to the potential energy of a neutral (since
$\sum_i s_i = 0$) plasma of quantized (since the $s_i$'s are
integers) charges $s_i$ on the lattice. The most relevant
$J([s_i])$ is therefore clearly the one that corresponds, in this
Coulomb analogy, to the lowest interaction energy. (Note that
strictly speaking $\Gamma$ corresponds to {\it twice} this energy,
because the sum in (\ref{Gammadef}) double counts). Aside from the
trivial configuration in which all the $s_i = 0$ , the lowest
energy configuration is clearly one in which there are two equal
and opposite unit magnitude charges on nearest neighbor sites:
i.e.,  $s_0 =+1  , {\vec \delta}_0 = {\vec 0}, s_1 =-1 , {\vec
\delta}_1 = {\vec a_\gamma}$. As discussed earlier, this
corresponds to the nearest-neighbor Josephson coupling in equation
(\ref{action1}). Thus, we have established that that coupling is,
indeed, the most relevant, as asserted earlier. Furthermore, using
simple symmetry arguments, one can show that for a symmetric
lattice (e.g., square, hexagonal, cubic), where all
nearest-neighbor sites are equivalent, $U({\vec a}_\gamma ) = -
\frac{1}{2 z_0}$, which recovers the recursion relation
(\ref{Vrg}) for $V$. Hence, all other couplings are {\it
irrelevant} for $\alpha \leq \alpha_c = 1/z_0$; as a result they
affect neither the floating phase nor the transition between it
and the ($D + 1$)-dimensionally coupled phase.

In summary, as long as $\alpha<1/z_0$, $V$, and all of the other
$J([s_i])$'s are   irrelevant. The ($0+1$)-dimensional systems
constitute locally critical power-law phases. Since $J$'s
are also irrelevant in this regime, the local systems are in a
floating phase \cite{Toner1}. As soon as $V$ becomes
relevant at $\alpha>1/z_0$, it grows to  $\infty$. In this limit,
this enormous periodic potential obviously traps the field $\Delta
\theta$ at the bottom of one of the minima of the periodic
potential; and provides a mass for fluctuations about that
minimum. Such a mass  destroys the logarithmically divergent
fluctuations, which give rise to the $\Gamma([s_i])/\alpha$ term
in the recursion relation in Eq.~\ref{Jrg} for $J$. Thus, in this
limit that recursion relation just becomes $dJ/dl = J$.  This in
turn makes all $J$'s relevant simultaneously with the nearest
neighbor coupling $V$, ensuring that there are no further
transitions as $\alpha$ is increased. Thus, there is only one
global transition with $\alpha$ as the tuning-parameter.

Even though the $J$'s  become relevant as soon as $V$ does, they
remain irrelevant far enough on the $\alpha$-axis for the
($0+1$)-dimensional transition to happen. The situation is
reversed for planes of power-law $XY$-phases with interlayer
couplings, and so Ref.~\cite{Toner1} had to introduce extra
gradient-couplings among the planes to get floating power-law
phases. In the present case, the metal to superconductor quantum
phase transition obtained by tuning dissipation in arbitrary
spatial dimension is then a true $(0+1)$-dimensional phase
transition. In the floating local critical phase on the metallic
side of the transition, the local systems are, however, not
decoupled; $J$'s are irrelevant, but not zero, and affect the
physical properties of the metallic phase only perturbatively.

To obtain higher order  corrections to the recursion
relation for $V$, Eq.~\ref{Vrg}, we perform a cumulant expansion
of Eq.~\ref{partition} in second order in $V$.
We
find no contribution to Eq.~\ref{Vrg} in $D=1$, while for $D=2$
triangular lattice there is a contribution at order $V^2$, which is
\begin{equation}
\frac{dV}{dl}=V(1-\frac{1}{z_0\alpha}) + C_1 V^2, \label{Vrg1}
\end{equation}
where $C_1$ is a positive constant \cite{future}. This holds for
any lattice in any dimension, as long as there are just three
sides in a minimum closed loop. For $2D$ square lattice, however,
there is no renormalization of $V$ at order $V^2$. Indeed,  the
first correction to Eq.~\ref{Vrg} comes only at order $V^3$
\cite{future} for any lattice where the minimum closed loop has
number of sides more than three,
\begin{equation}
\frac{dV}{dl}=V(1-\frac{1}{z_0\alpha}) + C_2 V^3, \label{Vrg2}
\end{equation}
where $C_2$ is another positive constant \cite{future}. It is important to
emphasize that the line of fixed points implied  by Eq. \ref{Vrg1} and
\ref{Vrg2} exist in any dimension, which implies that the upper critical
dimension is infinity.
Combined with the flow equation for $\alpha$,
$\frac{d\alpha}{dl}=0$, which remains true in our perturbative
treatment at all orders in $V$, we get the RG flows, the phase diagram, and
the exponents quoted earlier.

Not only the existence of a floating phase and its  validation  in the resistively shunted Josephson junction arrays are exciting possibilities, but the present analysis
should find its use in a much broader context: local quantum  critical points are
widely discussed in heavy fermion systems \cite{Si} and are also argued to
 be crucial to the ubiquitous metallic phases observed in a diverse class of
condensed matter systems  \cite{Mason}.

This work was supported by the NSF under grants:  DMR-0411931,
DMR-0132555 and DMR-0132726. We thank S. Kivelson and B. Spivak
for discussions, and  the Aspen Center for Physics for their
hospitality while a portion of this work was being completed.


\begin{thebibliography}{letter}

\bibitem{Toner1} C. S. O'Hern, T. C. Lubensky, and J. Toner, Phys.
Rev. Lett. \textbf{83}, 2745 (1999).

\bibitem{Emery}V. J. Emery {\em et al.}, Phys. Rev. Lett. {\bf 85},
2160 (2000).

\bibitem{DNA}C. S. O'Hern, T. C. Lubensky, Phys. Rev. Lett. {\bf 80},
4345 (1998); L. Golubovi{\'c} and M.
Golubovi{\'c}, Phys. Rev. Lett. {\bf 80}, 4341 (1998).

\bibitem{Safinya} T. Saldit, I. Koltover, J.O. R\"adler, and C. R.
Safinya, Phys. Rev. Lett. {\bf 79}, 2582
(1997); F. Artzner, R. Zantl, G. Rapp, and J.O. R\"adler, Phys.
Rev. Lett. {\bf 81}, 5015 (1998).

\bibitem{Goldman}See, for example, A. M. Goldman, Physica E {\bf 18},1 (2003).

\bibitem{Chakravarty0}S. Chakravarty {\em et al.}, Phys. Rev. Lett. {\bf
56}, 2003 (1986);

\bibitem{Fisherdiss}M. P. A. Fisher, Phys. Rev. B {\bf 36}, 1917 (1987).



\bibitem{CIKZ} S. Chakravarty {\em et al.} Phys. Rev. B {\bf 37}, 3283 (1988).

\bibitem{Fisher}M. P. A. Fisher, Phys. Rev. Lett. {\bf 57}, 585 (1986).

\bibitem{Larkin}M. V. Feigelman and A. I. Larkin, Chem. Phys. {\bf
235}, 107 (1998); B. Spivak, A. Zyuzin and M. Hruska, Phys. Rev. B {\bf 64}, 132502
(2001). These papers use an effective action different from ours. The description of the floating phase
remains unchanged, however,  but the line of unstable fixed
points in the ordered phase does not exist.

\bibitem{Takahide} Y. Takahide {\em et al.}, Phys. Rev. Lett. {\bf
85}, 1974 (2000); H. Miyazaki {\em et al.}
Phys. Rev. Lett. 89, 197001 (2002).

\bibitem{Kane}C. L. Kane and M. P. A. Fisher, Phys. Rev. B {\bf 46},
15233 (1992).

\bibitem{Kadin}A. M. Kadin, K. Epstein, and A. M. Goldman, Phys. Rev.
B 27, 6691-6702 (1983).

\bibitem{nodub}We exclude ``loops'' of two bonds made by simply retracing
the same bond twice in opposite directions.

\bibitem{Caldeira} A. O. Caldeira and A. J. Leggett, Ann. Phys.
\textbf{149}, 374 (1983).



\bibitem{Schmid2}A. Schmid, in {\em The art of measurement:
metrology in fundamental and applied physics}, edited by B. Kramer
(VCH Verlagsgesellschaft mbH, Weinheim, 1988)


\bibitem{dangerous}Some of the correlation functions  depend on
  $1/E_0$   singularly; to calculate these we must keep a non-zero
  $1/E_0$ . In this sense  $1/E_0$ is a ``dangerously irrelevant" variable.




\bibitem{rough}See, e.g., S. T. Chui and J. D. Weeks, Phys. Rev.
B{\bf 14}, 4978 (1976).

\bibitem{srough} J. Toner and D. P. DiVincenzo, Phys. Rev. B {\bf
41}, 632 (1990).

\bibitem{future}S. Tewari, J. Toner, and S. Chakravarty, in preparation.




\bibitem{Si}Q. M. Si {\em et al.}, Nature  {\bf 413}, 804 (2001).

\bibitem{Mason}A. Kapitulnik {\em et al.}, Phys. Rev. B {\bf 63},
125322 (2001).


\end{thebibliography}
\end{document}